\documentclass[12pt]{elsarticle}

\biboptions{sort&compress}
\usepackage[T1]{fontenc}
\usepackage[utf8]{inputenc}
\usepackage{microtype}
\usepackage{amsmath,amssymb}
\usepackage{siunitx}
\usepackage[colorlinks=true,
            linkcolor=blue,
            citecolor=blue,
            urlcolor=blue,
            filecolor=blue]{hyperref}
\usepackage{doi}
\usepackage{graphicx}
\usepackage{booktabs}
\usepackage{tabularx}
\usepackage{subcaption}
\usepackage{float}
\usepackage[section]{placeins}

\usepackage{cleveref}
\crefname{figure}{Fig.}{Figs.}
\Crefname{figure}{Fig.}{Figs.}
\crefname{table}{Table}{Tables}
\Crefname{table}{Table}{Tables}
\usepackage{threeparttable}
\usepackage{tikz}
\usetikzlibrary{arrows.meta,calc,fit,positioning,shapes.geometric}

\newcommand{\chg}[1]{#1}
\newcommand{\fabnote}[1]{}
\newcommand{\rmv}[1]{}

\makeatletter
\def\ps@pprintTitle{%
  \let\@oddhead\@empty
  \let\@evenhead\@empty
  \let\@oddfoot\@empty
  \let\@evenfoot\@oddfoot}
\makeatother

\begin{document}

\begin{frontmatter}

\title{\vspace*{-20mm}Design under Throughput Scale-Up: Equations that Transfer the
       Pressure--Conversion Front to Untested Flow Rates in a Reactive
       Tesla-Valve Micromixer}

\address[ntnu]{Department of Chemical Engineering, Norwegian University of
               Science and Technology (NTNU), Trondheim, Norway}
\address[imperial]{Department of Chemical Engineering, Imperial College London,
                   London, UK}
\address[surrey]{Surrey Institute for People-Centred Artificial Intelligence,
                 University of Surrey, Guildford, UK}
\address[cimatec]{SENAI CIMATEC University, Salvador, Brazil}
\address[puffin]{Puffin Hub, Trondheim, Norway}

\author[ntnu]{Patrick Souza Lima}
\author[imperial]{Ehecatl Antonio del R{\'i}o Chanona}
\author[surrey,cimatec]{Erick G. Sperandio Nascimento}
\author[ntnu,puffin]{Idelfonso B. R. Nogueira\corref{cor1}}
\cortext[cor1]{Corresponding author. Email: idelfonso.b.d.r.nogueira@ntnu.no}

\begin{abstract}
\footnotesize
{Raising the flow rate through a reactive micromixer shortens residence
time and raises pressure drop, so a geometry chosen at one flow rate may not
suit another, and repeating a CFD optimisation at every flow rate is costly.
We test whether short equations fitted across optimisation campaigns recover
the pressure--conversion front at a flow rate that was not optimised. A
Tesla-valve micromixer running the Bourne neutralisation--hydrolysis reactions
was optimised by multi-fidelity Bayesian optimisation at seven velocities over
an eightfold range. Symbolic regression of the seven surrogates, with no
functional form imposed, returned the same equations in three independent
searches: pressure drop is a hydraulic resistance proportional to flow rate,
and the effective Damk\"ohler number is set by a channel length divided by
flow rate, a contact time. A diffusion-only estimate reproduces the fine-mesh
conversion, so in this two-dimensional laminar model the stages act through
contact time rather than added mixing, and the efficient geometries change
little with flow rate. Equations of 12 nodes found without one flow rate
placed the front there within 0.5 percentage points of conversion and kept
95\% of its hypervolume; longer equations fitted better without predicting
better fronts. Twenty new fine-mesh designs taken from predicted fronts
confirmed the accuracy of the equations and the size of that shortfall.
Inside the range already sampled, a fine-mesh check of a few designs read
from the equations, about two hours of CFD, replaces a new optimisation of
about 40 hours. For the Tesla
valve, the highest conversion fell by half across the flow range, and keeping
conversion while doubling the flow rate per channel requires twice the device
length and about four times the pressure drop.}
\end{abstract}


\end{frontmatter}

\section{Introduction}
\label{sec:introduction}

Microfluidic systems enable precise control of transport and reaction in
chemical manufacturing, pharmaceutical processing, diagnostics, and other
intensified operations \citep{Cai2023,Raza2020,Ganguli2024}. Their small
dimensions, however, make mixing strongly diffusion-limited under the
low-Reynolds-number conditions typical of microdevices. Passive micromixers address
this limitation through geometry alone, but features that increase interfacial
stretching and transverse transport commonly also increase pressure drop
\citep{Ganguli2024,Raza2020,Su2014}. \chg{Design therefore means choosing a
point on the trade-off between pressure drop and reactive performance.}

That decision changes when throughput is increased{: raising} inlet velocity
shortens residence time while increasing hydraulic demand, so a design chosen
at one operating condition may violate a conversion target or pressure limit
at another. Repeating a complete CFD optimisation for every flow rate is
costly and can return disconnected sets of preferred geometries without
revealing how the underlying decision changes. A useful scale-up result must
instead connect operating condition, geometry, pressure, and conversion well
enough to screen a design for a stated process requirement
{\citep{Nogueira2019SPSO,Nogueira2022NestedLoop}}.

\chg{Microreactor throughput is usually raised by numbering-up, which keeps each
channel at its design point but requires even flow distribution over many
units, or by sizing-up, for example by lengthening the channel
\citep{Dong2021}.} {This paper measures the cost of the second route for a
single channel, so that it can be compared with the first: the flow rate is
raised and the device is lengthened as needed, which reduces the number of
parallel units at a cost in conversion and pressure drop.} \chg{Here} {throughput} \chg{is the inlet volumetric flow
rate per unit depth of the two-dimensional model. Channel width and the base
cell are fixed; the stage count, pitch, skew, and loop proportions, and
therefore the device length, are re-optimised at each flow rate.}

Tesla valve-based micromixers \chg{suit this problem because their geometry sets
both the hydraulic cost and the mixing performance.} Their
asymmetric channels and repeated splitting, recombination, and local
recirculation create multiple geometric routes to enhance scalar transport
without external actuation {\citep{Abolpour2022,Guo2022,Sun2024,Li2025,Liu2025,
Yang2025,Juraeva2022Tesla}}. {Whether these routes add to reactive
conversion at the low Reynolds numbers considered here, beyond the residence
time that a longer channel provides, is part of what this case tests.}
{Their geometry has been designed by
parametric optimisation, simulated annealing, and topology optimisation
\citep{Abolpour2022,WangChen2022SA,Xiong2023,Li2025,LiTesla2026}. These
studies optimise at one operating condition, sometimes checking the optimum
at other flow rates afterwards, or compare fixed geometries over a range of
Reynolds numbers. None re-optimises the geometry as the flow rate changes.
Outside Tesla designs, a serpentine mixer has been optimised separately at a
low and a high Reynolds number \citep{Raza2018}, with no rule linking the two
optima. Similarly, a toroidal micromixer has been optimised by multi-objective
Bayesian optimisation at three inlet flow-rate ratios, with the total flow
rate fixed, and the change of the optimal geometry between conditions could
not be identified beyond the inlet angle \citep{Fonseca2025BayesShape}.}

{Judging a geometry at several flow rates also needs a reactive measure.}
\chg{The Bourne neutralisation--hydrolysis
system supplies {it} \citep{Baldyga1999,Bourne2003}. Its
neutralisation step is effectively instantaneous, so the HCl conversion at the
outlet measures how much of the acid stream the base has reached, and it falls
as residence time shortens. The Tesla--Bourne system is used here as a benchmark in
which geometry, pressure drop, and reaction performance must be judged together
as the flow rate changes.}

\chg{CFD is increasingly used in process systems engineering as a source of
decisions as well as of flow fields, for example to identify transitional
multiphase flow regimes \citep{Lima2026FuzzyCFD} and to design a stabilising
control structure \citep{Lima2025CAFControl}. Choosing a micromixer geometry
for each throughput is a decision of the same kind, but it requires} many
reactive CFD evaluations across a mixed continuous--discrete geometry family and
several \chg{flow rates}. Dense
parameter sweeps are impractical, whereas Bayesian optimisation can direct
new evaluations toward informative parts of an expensive black-box design
space {\citep{Shahriari2016,Frazier2018,Garnett2023,Kundacina2025,
Basha2023CEJ,Savage2024NatChemEng}}.
\chg{It is derivative-free and sample-efficient, which suits evaluations that
depend on commercial simulators or otherwise costly models}
\citep{Zhao2026,Lima2026Burner}.
{Treating mesh size as a continuous fidelity variable lets coarse meshes
explore the geometry while fine meshes are reserved for candidate designs
\citep{Savage2023}; in earlier CFD design studies this mesh-fidelity structure
shortened the optimisation campaigns
\citep{Lima2026Burner,Lima2026BurnerMFBO}.}

{Each campaign, although cheaper than a single-fidelity one, still returns
designs only for the conditions it explored. Where a flow condition is itself a decision variable
\citep{Savage2023}, the optimisation returns one best condition with its
design. Multi-task and contextual Bayesian optimisation reuse earlier
campaigns to speed up later ones \citep{Swersky2013,Krause2011}, and
parametric multi-objective methods learn the Pareto set as a function of a
parameter \citep{WeaverRosen2023,Cheng2026PPSLMOBO}. Both return an optimiser
or a learned model; neither gives a relation that an engineer can read and
evaluate.}

\chg{The CFD population produced by the optimisation is useful beyond locating
the best designs at each flow rate}{:} interpretable models can
compress the observed \chg{design--flow-rate} relationships into screening rules.
{Multiparametric programming gives explicit solutions as functions of
parameters, but requires an algebraic model \citep{Oberdieck2016}, which a CFD
simulation does not provide; algebraic surrogate modelling and symbolic
regression can supply one from simulation data
{\citep{Cozad2014,Wilson2017,Engle2022,Rebello2022MLGuide}}.}
{Hybrid models that embed a neural network in the governing equations can
fit such data with little loss of physical structure
\citep{Nogueira2022UDE,Lima2023UDECryst}, but the learned term remains a black
box.} Symbolic regression is suited to this {task} because it produces explicit
relations that can be inspected and tested
{\citep{Ansari2022,Rogers2024,Rebello2024SciML}}. Free-form evolutionary
methods search mathematical structures directly \citep{Schmidt2009,
Cranmer2023PySR}; sparse-library methods expose the accuracy--complexity
trade-off within a declared function set \citep{Brunton2016}.

{We found no study that compresses the CFD data of several optimisations
into short equations that predict the trade-off front at a flow rate that was
never optimised, and then tests designs read from those equations in new
fine-mesh simulations. This paper does so for a five-variable Tesla-valve
family operated at seven flow rates. Each optimisation surrogate
describes one flow rate; an equation fitted to all of them carries the flow
rate as a variable and can be evaluated where no optimisation was run.}

{The objectives are to (i) obtain the pressure--conversion trade-off front at each
flow rate on one common fine mesh; {(ii) derive, by free-form symbolic
regression of the campaign surrogates, short equations for pressure drop and
conversion in terms of flow rate and geometry, and read their physical
meaning; (iii) test whether equations found without one flow rate recover the
front at that flow rate, first against the optimised front and then with new
fine-mesh simulations; and
(iv) use the equations to draw conclusions on the device: what controls its
conversion, and what raising the flow rate per channel costs in length and
pressure drop.}}

\section{Methodology}
\label{sec:methodology}

\begin{figure}[H]
\centering
\resizebox{\textwidth}{!}{\input{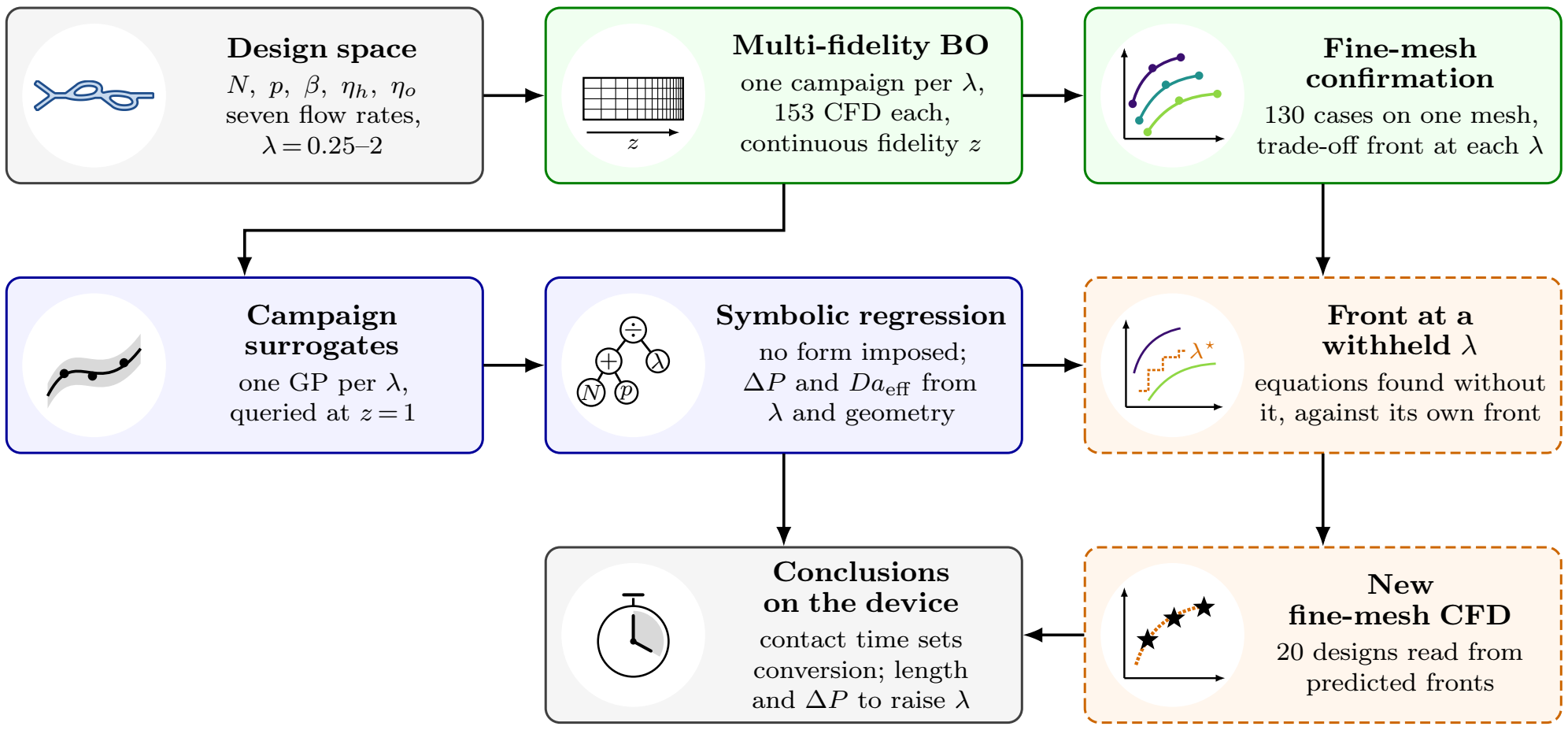}}
\caption{Workflow of the study. Green, CFD; blue, surrogates and equations;
orange with dashed border, tests at a withheld flow rate.}
\label{fig:methodology_workflow}
\end{figure}

{\Cref{fig:methodology_workflow} shows the study in two parts. In the
first, a separate multi-fidelity Bayesian optimisation is run at each of seven
flow rates, and the designs it selects are recomputed on one fine mesh; this
gives the observed trade-off front at each flow rate. In the second, the seven
campaign surrogates are condensed by symbolic regression into equations in
which the flow rate is a variable, and the equations are tested at flow rates
withheld from the search, first against the optimised fronts and then with new
fine-mesh simulations.}

\FloatBarrier

\subsection{{Design space, CFD model, and objectives}}
\label{subsec:workflow}
\label{subsec:cfd}

\begin{figure}[H]
\centering
\includegraphics[width=\textwidth]{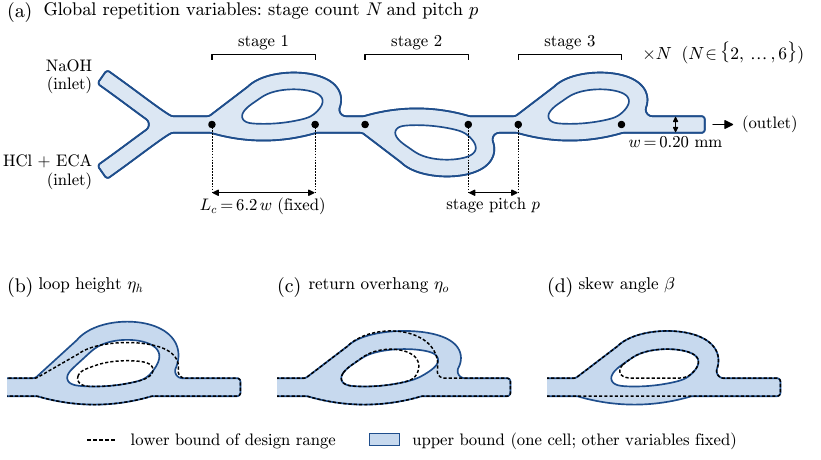}
\caption{{Geometric design variables. (a) $N$ and $p$, drawn at
$p=\SI{0.60}{mm}$; dots, ends of each stage. (b)--(d) $\eta_h$, $\eta_o$,
and $\beta$ at the two bounds of their ranges, at one physical scale.}}
\label{fig:design_variables}
\end{figure}

{Each CFD evaluation returns pressure drop and HCl conversion for a traced
Tesla architecture. The channel width $w=\SI{0.20}{mm}$, the base cell, and
the base inlet velocity $u_0=\SI{0.02}{m.s^{-1}}$ are fixed; the stage count
and the cell shape are re-optimised at each flow rate, so the device length
changes with them.}
The micromixer topology consists of two inlets, a repeated traced Tesla-valve
section, and a common outlet.
The geometry generator constructs the main channel and return-loop structures
from centreline paths and offsets them by the channel width.
The physical geometry uses a mixed continuous--discrete design space,
\begin{equation}
  \mathbf{x}_g =
  \left[
  p,\ \beta,\ \eta_h,\ \eta_o,\ N
  \right],
  \label{eq:design-vector}
\end{equation}
where $p$ is the stage pitch, $\beta$ is the stage skew angle, $\eta_h$ is the
loop-height factor, $\eta_o$ is the return-overhang factor, $N$ is the number
of Tesla stages.
The four architectural variables are continuous, whereas $N$ is discrete.
Within a given geometry, all Tesla stages are generated from the same repeated
cell definition; therefore $p$, $\beta$, $\eta_h$, and $\eta_o$ are constant
from stage to stage.
The optimisation changes the size and shape of this repeated stage between
candidate geometries, while $N$ changes how many identical stages are placed in
series.
\Cref{fig:design_variables} illustrates each geometric variable, and
\cref{tab:design_space} gives its range{.}

\begin{table}[H]
\centering
\caption{{Design variables and their ranges.}}
\label{tab:design_space}
\small
\renewcommand{\arraystretch}{1.12}
\begin{tabular}{@{}lll@{}}
\toprule
\textbf{Variable} & \textbf{Range} & \textbf{Type} \\
\midrule
Stage pitch, $p$ & 0.00--\SI{0.60}{mm} & Continuous \\
Stage skew angle, $\beta$ & 0.0--\SI{36.0}{\degree} & Continuous \\
Loop-height factor, $\eta_h$ & 2.30--3.65 & Continuous \\
Return-overhang factor, $\eta_o$ & 0.30--0.82 & Continuous \\
Number of stages, $N$ & $\{2,3,4,5,6\}$ & Discrete \\
\bottomrule
\end{tabular}
\end{table}

The physical model, material properties, and boundary conditions are collected
in \cref{tab:cfd-conditions}{. Over the whole sweep the flow is laminar
and mixing is transport limited.}

\begin{table}[H]
\centering
\caption{{Physical model, material properties, inlet compositions, and
boundary conditions. For a given geometry, $\lambda$ changes only the inlet
velocity.}}
\label{tab:cfd-conditions}
\small
\renewcommand{\arraystretch}{1.12}
\begin{tabularx}{\textwidth}{@{}p{0.31\textwidth}X@{}}
\toprule
\textbf{Item} & \textbf{Specification} \\
\midrule
Flow and transport model & Two-dimensional, steady, laminar, isothermal flow
with species transport; gravity neglected \\
Temperature & \SI{298}{K}; energy equation not solved \\
Mixture density & $\rho=\SI{998.2}{kg.m^{-3}}${, constant} \\
Dynamic viscosity & $\mu=\SI{1.003e-3}{kg.m^{-1}.s^{-1}}$ \\
Species mass diffusivity & $D=\SI{1.0e-9}{m^2.s^{-1}}$ \\
Reference channel width & $w=\SI{0.20}{mm}$ \\
Velocity inlets & $u_{\mathrm{in}}=\lambda u_0$, with
$u_0=\SI{0.02}{m.s^{-1}}$ and
$\lambda\in\{0.25,0.35,0.5,0.7,1,1.5,2\}$, on flat inlet segments of
length $\SI{0.124}{mm}$ \\
Pressure outlet & Zero gauge pressure \\
Upper-inlet composition &
$C_{\mathrm{NaOH},0}=\SI{90}{mol.m^{-3}}$ \\
Lower-inlet composition &
$C_{\mathrm{HCl},0}=\SI{45}{mol.m^{-3}}$ and
$C_{\mathrm{ECA},0}=\SI{90}{mol.m^{-3}}$ \\
Reynolds number & $Re=\rho u_{\mathrm{in}}w/\mu\simeq1$--8 \\
P\'eclet number & $Pe=u_{\mathrm{in}}w/D\simeq1.0\times10^3$--$8.0\times10^3$ \\
\bottomrule
\end{tabularx}
\end{table}

Reactive performance is evaluated {with} the Bourne/ECA chemistry
{implemented} in a Fluent user-defined function.
As listed in \cref{tab:cfd-conditions}, the upper stream contains sodium
hydroxide, while the lower stream contains hydrochloric acid and ethyl
chloroacetate (ECA).
The source terms include the fast neutralisation reaction
\begin{equation}
  \mathrm{NaOH}+\mathrm{HCl} \rightarrow \mathrm{NaCl}+\mathrm{H_2O},
  \qquad
  k_1 = \SI{1.3e8}{m^3.mol^{-1}.s^{-1}},
  \label{eq:neutralisation}
\end{equation}
and the slower ECA hydrolysis pathway
\begin{equation}
  \mathrm{NaOH}+\mathrm{ECA} \rightarrow \mathrm{SCA}+\mathrm{C_2H_5OH},
  \qquad
  k_2 = \SI{3.0e-2}{m^3.mol^{-1}.s^{-1}},
  \label{eq:hydrolysis}
\end{equation}
{where SCA is sodium chloroacetate. The hydrolysis competes with the
neutralisation for NaOH; designs are compared by HCl conversion.}

{The flow and species fields satisfy the steady incompressible equations
\begin{align}
  \nabla\cdot\mathbf{u} &= 0, \label{eq:continuity}\\
  \rho\,(\mathbf{u}\cdot\nabla)\mathbf{u} &= -\nabla p + \mu\nabla^{2}\mathbf{u},
  \label{eq:momentum}\\
  \rho\,\mathbf{u}\cdot\nabla Y_i &= \nabla\cdot\left(\rho D\nabla Y_i\right)
  + M_i\textstyle\sum_{j}\nu_{ij}r_j, \label{eq:species}
\end{align}
where $Y_i$, $M_i$, and $\nu_{ij}$ are the mass fraction, molar mass, and
stoichiometric coefficient of species $i$ in reaction $j$. Both reactions are
second order, with rates
\begin{equation}
  r_1 = k_1 C_{\mathrm{NaOH}}C_{\mathrm{HCl}}, \qquad
  r_2 = k_2 C_{\mathrm{NaOH}}C_{\mathrm{ECA}}, \qquad
  C_i = \rho Y_i/M_i,
  \label{eq:rates}
\end{equation}
so NaOH is consumed by $r_1+r_2$, HCl by $r_1$, ECA by $r_2$, and SCA is
formed by $r_2$; NaCl, ethanol, and water are lumped into the solvent. The
rates are implemented as source terms in a user-defined function. Because
$k_1C$ is of order \SI{e9}{s^{-1}}, far faster than any transport in the
device, NaOH and HCl react wherever they meet: the reaction is confined to a
thin layer along the interface between the two streams, and its rate is set
by the supply of reactants to that layer.}

The optimisation problem is formulated as a two-objective minimisation problem.
The hydraulic objective is the pressure drop between the two inlet surfaces and
the outlet,
\begin{equation}
  J_1(\mathbf{x}_g) = \Delta P =
  \bar{P}_{\mathrm{inlets}}-\bar{P}_{\mathrm{outlet}},
  \label{eq:pressure-drop}
\end{equation}
where the pressures are area-weighted surface averages.
The reactive objective is the unconverted HCl fraction,
\begin{equation}
  J_2(\mathbf{x}_g) = 1-X_{\mathrm{HCl}} =
  \frac{\dot{n}_{\mathrm{HCl,out}}}
       {\dot{n}_{\mathrm{HCl,in}}},
  \label{eq:hcl-unconverted}
\end{equation}
where $X_{\mathrm{HCl}}$ is the HCl conversion and
$\dot{n}_{\mathrm{HCl,in}}$ and $\dot{n}_{\mathrm{HCl,out}}$ are the inlet and
outlet HCl molar flow rates.
The target is therefore
\begin{equation}
  \min_{\mathbf{x}_g} \left[J_1(\mathbf{x}_g), J_2(\mathbf{x}_g)\right]{,}
  \label{eq:multiobjective-problem}
\end{equation}
{and each reported front is the set of observed designs for which
conversion cannot be raised without raising pressure drop
\citep{Miettinen1999}.}
To separate single-pass conversion from the amount converted as inlet flow is
varied, a dimensionless relative converted HCl flow is also defined as
\begin{equation}
  \Phi_X=\lambda X_{\mathrm{HCl}}
  =\frac{\dot n_{\mathrm{HCl,converted}}(\lambda)}
  {\dot n_{\mathrm{HCl,in}}(\lambda=1)}{,}
  \label{eq:converted-throughput}
\end{equation}
{since the inlet molar flow is proportional to $\lambda$. It is a
post-processing quantity, not an objective.}

{For the equations of \cref{subsec:symbolic-method}, conversion is
expressed as an effective Damk\"ohler number,}
\begin{equation}
  Da_{\mathrm{eff}}=-\ln(1-X_{\mathrm{HCl}}),
  \label{eq:effective-da}
\end{equation}
{the Damk\"ohler number that an ideal first-order plug-flow reactor would
need to reach the same conversion. Because the neutralisation is effectively
instantaneous, $Da_{\mathrm{eff}}$ measures mixing rather than kinetics. It
keeps the back-transformed conversion between zero and one, it is close to a
power of $\lambda$, and that power has two reference values: $-1$ when the
reaction is kinetically controlled, since $Da$ is then proportional to
residence time, and $-1/2$ when two streams react only by diffusion across
their interface, since the reacted layer then grows as the square root of
the contact time.}

\subsection{{Optimisation at each flow rate and fine-mesh confirmation}}
\label{subsec:bomfmo}
\label{subsec:fidelity}

{Unstructured meshes refined at the walls are generated with Gmsh
\citep{Geuzaine2009}.}
The numerical fidelity is controlled by the characteristic mesh size $s$.
Rather than treating fidelity as a small number of categorical mesh levels, the
campaign uses a continuous coordinate $z\in[0,1]$,
\begin{equation}
  z =
  \frac{1/s - 1/s_{\max}}
       {1/s_{\min} - 1/s_{\max}},
  \qquad
  s_{\min}=\SI{0.005}{mm},\quad
  s_{\max}=\SI{0.020}{mm}.
  \label{eq:fidelity-transform}
\end{equation}
Thus, $z=0$ corresponds to the coarsest mesh and $z=1$ to the finest mesh used
as the optimisation target.

\begin{figure}[H]
\centering
\includegraphics[width=0.9\textwidth]{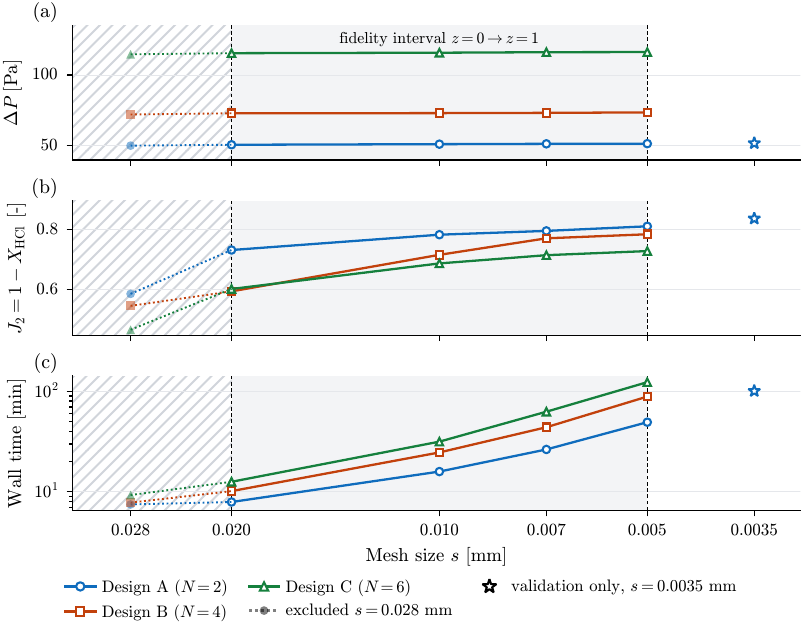}
\caption{{Mesh response at $\lambda=2$ for three representative
geometries: (a) pressure drop, (b) unconverted HCl, and (c) wall time. Dashed lines and shading mark the
fidelity interval from $z=0$ to $z=1$; open stars, a finer validation mesh
($s=\SI{0.0035}{mm}$); translucent markers in the hatched band, an excluded
coarser level ($s=\SI{0.028}{mm}$).}}
\label{fig:mesh_response}
\end{figure}

{The fine mesh was chosen from a mesh-response study of three
representative geometries at the highest flow rate, $\lambda=2$, where the
mesh matters most, with a check at $\lambda=0.25$
(\cref{fig:mesh_response}). Pressure drop is nearly insensitive to the mesh.
Conversion is more sensitive, as expected for a fast reaction that takes place
in a thin layer between the two streams. The mesh $s=\SI{0.005}{mm}$
was adopted because refining it further changes conversion by only a few
percentage points while doubling the cost (\cref{fig:mesh_response}). This
remaining change grows with flow rate, from about half a percentage point
at $\lambda=0.25$ to 1.4--2.6 points at $\lambda=2$, so comparisons between
flow rates carry more numerical uncertainty than comparisons at one flow
rate. HCl
conversion is therefore read as an outlet-integrated response on this mesh,
and every comparison below is made on it.}

\FloatBarrier

{Every case is solved with ANSYS Fluent 2024 R2 in double precision, with
the pressure-based coupled algorithm and second-order discretisation, and is
accepted once pressure drop and HCl conversion stop changing and the mass and
species balances close.}

{Each flow rate has its own campaign. A nested design of experiments (DOE) of 30
feasible geometries at $z=0$, with subsets of 15 at $z=1/3$ and 8 at $z=1$
(53 evaluations, Latin hypercube for the continuous variables
\citep{McKay1979}), is followed by 100 sequential evaluations, for 153 per
flow rate and 1071 in total. Separate Gaussian processes with automatic
relevance determination model $\log\Delta P$, $\log(1-X_{\mathrm{HCl}})$, and
wall time on the input $[\mathbf{x}_g,z]$ \citep{Rasmussen2006}; $N$ is
queried only at its integer values. At each step the acquisition scores 5000
feasible candidates by
\begin{align}
  A(\mathbf{x}_g,z)&=
  \frac{\Delta HV_{z=1}(\mathbf{x}_g)\,R_{z\to1}(\mathbf{x}_g)}
       {\hat t(\mathbf{x}_g,z)},
  \label{eq:acquisition}\\[4pt]
  R_{z\to1}(\mathbf{x}_g)&=\frac{\operatorname{Cov}[f(\mathbf{x}_g,z),f(\mathbf{x}_g,1)]^2}
  {\operatorname{Var}[f(\mathbf{x}_g,z)]\,\operatorname{Var}[f(\mathbf{x}_g,1)]},
  \label{eq:variance-reduction}
\end{align}
where $\Delta HV_{z=1}$ is the hypervolume improvement \citep{Zitzler1999}
obtained when the lower confidence bounds of the two objectives predicted at
$z=1$ are added to the current front, $R_{z\to1}$ is the
fraction of the variance at $z=1$ that an evaluation at $z$ would remove,
taken from the GP covariance along $z$ and averaged over the two objectives,
and $\hat t$ is the predicted wall time. A cheap evaluation is therefore
chosen only when it is informative about the fine mesh.}

{After each campaign, the
evaluated geometries are projected to $z=1$ through the surrogates, and ten
designs spread along the projected front, including its two ends and its
knee, are recomputed on the fine mesh. With the eight $z=1$ designs of the DOE,
this gives 18 to 22 fine-mesh cases per flow rate. Because the error of a
coarse mesh is problem dependent \citep{Mikkola2023}, every front reported
below is the nondominated set of these observed fine-mesh responses, never of
surrogate predictions.}
\label{subsec:target-selection}

\subsection{{Design equations and the withheld-flow-rate test}}
\label{subsec:symbolic-method}

{The equations are fitted to the campaign surrogates rather than to the
CFD cases, so that every region of the design space is represented at every
flow rate. Each of the seven Gaussian processes is queried at $z=1$ on 5000
Latin-hypercube designs per flow rate, balanced over $N$; queries with a
large posterior uncertainty are discarded, which leaves 30\,904 designs, and
400 of them are drawn at random per flow rate. Symbolic regression then
searches over mathematical expressions instead of adjusting the coefficients
of a form chosen beforehand \citep{Schmidt2009,Cranmer2023PySR}; the search
was run with PySR, and no functional form was supplied. Pressure drop and
$Da_{\mathrm{eff}}$ (\cref{eq:effective-da}) are searched separately, with
the squared error weighted by $1/y^2$ so that small and large responses count
alike, and \cref{tab:sr-setup} lists the settings.}

{The length of an expression, its complexity, is the number of
variables, constants, and operators it contains, the nodes of its
expression tree. For each
complexity the search returns the most accurate expression it found. The
complexity at which a variable first appears shows how much of the response
that variable explains, and the list shows where more length stops paying.
The search is stochastic, so it was run three times with different random
seeds on all seven flow rates, and only expressions found in all three runs
are interpreted.}

{Which length to use is decided by how well an equation predicts a flow
rate it has not seen, not by how well it fits. One flow rate, $\lambda^\star$,
is withheld, and the search is repeated with the other six. The resulting
equations are evaluated at $\lambda^\star$ on $2^{14}$ Sobol designs for each
$N=2,\dots,6$, kept inside the region covered by the six training
surrogates, and their nondominated designs form the predicted front; no
response at $\lambda^\star$ is used to build it. The campaign surrogate of
$\lambda^\star$, which took no part in the search, then scores this front in
two ways: the conversion lost at equal pressure drop relative to its own
front, averaged over 25 pressure levels, and the fraction of its hypervolume
retained. This is repeated for every $\lambda^\star$ and every complexity
limit, and the working equations are the shortest whose loss is within 0.15
percentage points of the lowest loss over all limits. Because the scoring
uses a surrogate rather than CFD, its agreement with fine-mesh CFD is
reported in {\cref{subsec:results-symbolic}}.}

{Fine-mesh CFD tests the equations in two further ways. First, the
equations found without $\lambda^\star$ predict the 18 to 22 fine-mesh cases
that the campaign at $\lambda^\star$ already contains. Second, twenty designs
taken from predicted fronts were simulated on the fine mesh after the
prediction was made: twelve at $\lambda^\star=0.35$, 1, and 2, from an
earlier fit of fixed power-law form, and eight at $\lambda^\star=0.5$, 1, and
1.5, from the fronts of the working
equations.}

\begin{table}[H]
\centering
\caption{{Symbolic search and withheld-flow-rate test.}}
\label{tab:sr-setup}
\small
\begin{tabular*}{\textwidth}{@{}ll@{\extracolsep{\fill}}ll@{}}
\toprule
\multicolumn{2}{@{}l}{\textbf{Search}} &
\multicolumn{2}{@{}l@{}}{\textbf{Withheld-flow-rate test}} \\
\cmidrule(r{1.25em}){1-2}\cmidrule{3-4}
Inputs & $\lambda$, $N$, $p/w$, $\beta$, $\eta_h$, $\eta_o$ &
Designs evaluated & 81\,920 \\
Operators & $+$, $-$, $\times$, $\div$, power, exp &
Pressure levels & 25 \\
Populations & 24 of 40 expressions &
Complexity limits & 6, 8, \dots, 30 \\
Iterations & 2000 &
Tolerance & 0.15 points \\
Maximum complexity & 30 & & \\
\bottomrule
\end{tabular*}
\end{table}

\section{Results and Discussion}
\label{sec:results}

{\Cref{subsec:results-mesh} shows how the seven searches converged and
gives the observed trade-off fronts at the seven flow rates. \Cref{subsec:results-complexity} presents the design
equations {and the physics they express. \Cref{subsec:results-symbolic}
tests them at flow rates that were not optimised, fixes how long they need to
be, checks them with new fine-mesh simulations, and uses them to price each
design lever.}}

\subsection{{Trade-off fronts at seven flow rates}}
\label{subsec:results-mesh}

\label{subsec:results-allocation}
\label{subsec:results-decision-map}

\begin{figure}[H]
\centering
\includegraphics[width=\textwidth]{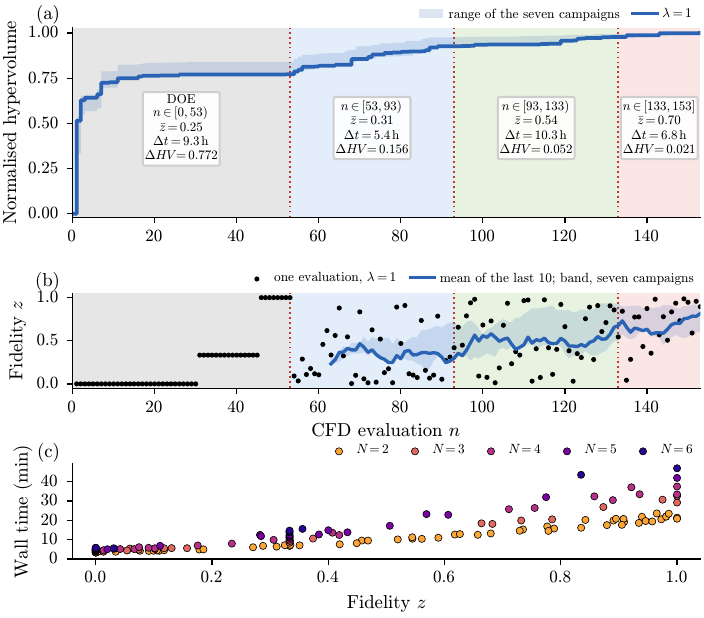}
\caption{{Optimisation at $\lambda=1$; bands, range of the seven
campaigns. (a) Hypervolume at $z=1$ predicted by the final surrogate,
normalised; boxes, mean fidelity, wall time, and hypervolume gain per stage. (b) Fidelity of each evaluation; line, mean of
the last ten. (c) Wall time against fidelity; colour, $N$.}}
\label{fig:hv-stages}
\end{figure}

{All seven campaigns completed their 153 evaluations, 1071 in total, and
every adaptive evaluation used an intermediate fidelity ($0<z<1$, mean
0.55). \Cref{fig:hv-stages}a shows the pattern expected from a space-filling
start followed by a guided search.}

{The hypervolume rises quickly over the
first DOE evaluations and then levels off, at 0.77 to 0.84 of its final value
across the seven campaigns: once the design space is covered, further
spread-out designs rarely extend the front.}

\begin{figure}[H]
\centering
\includegraphics[width=\textwidth]{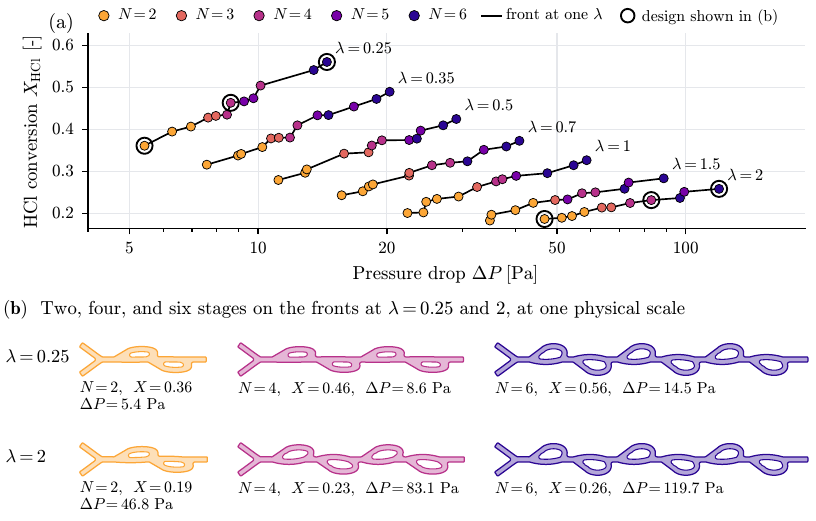}
\caption{{Observed fronts on the fine mesh. (a) HCl conversion against
pressure drop, one front per flow rate; colour, number of stages $N$;
rings, the designs of (b). (b) Designs with two, four, and six stages on the fronts at
$\lambda=0.25$ and 2, at one physical scale.}}
\label{fig:common-resolution-map}
\end{figure}

{The adaptive evaluations resume
the growth because they are placed where the front can still move. Once the
coarse meshes have told the surrogates what they can about the fine mesh,
only evaluations close to $z=1$ remain informative (\cref{eq:acquisition}),
and the mean fidelity of the adaptive evaluations rises
(\cref{fig:hv-stages}b). Each evaluation therefore costs more, since the wall
time grows steeply with fidelity and with the number of stages
(\cref{fig:hv-stages}c), while adding less; in the last stage the gain
per hour of CFD is about a tenth of that in the first adaptive stage.}

{All seven campaigns follow this pattern and were given the same budget of 100
adaptive evaluations, so that every flow rate received the same search
effort. After the search, the designs on each predicted front were
recomputed on the fine mesh. Of the 83 designs that form the observed fronts,
64 came from the optimisation and 19 from the DOE, in line with the
hypervolume gain of \cref{fig:hv-stages}.}

\FloatBarrier

\FloatBarrier

{Each observed front is the nondominated set of the fine-mesh cases at
its flow rate (\cref{fig:common-resolution-map}a). Per pass, a lower flow rate is better on both objectives: the front at
$\lambda=0.25$ lies entirely above and to the left of the one at $\lambda=2$
(\cref{fig:common-resolution-map}a).}

{A slower stream gives the reactants more time in contact, so it converts
more, and in the nearly creeping flow here ($Re=1$--8) it also costs less,
because the pressure drop grows almost linearly with velocity. Along each
front, conversion is raised mainly by adding stages
(\cref{fig:common-resolution-map}b), which also lengthens the contact time,
at a price in pressure drop; at a higher velocity each stage adds less
contact time and costs more, so the fronts flatten.}

\begin{figure}[H]
\centering
\includegraphics[width=0.92\textwidth]{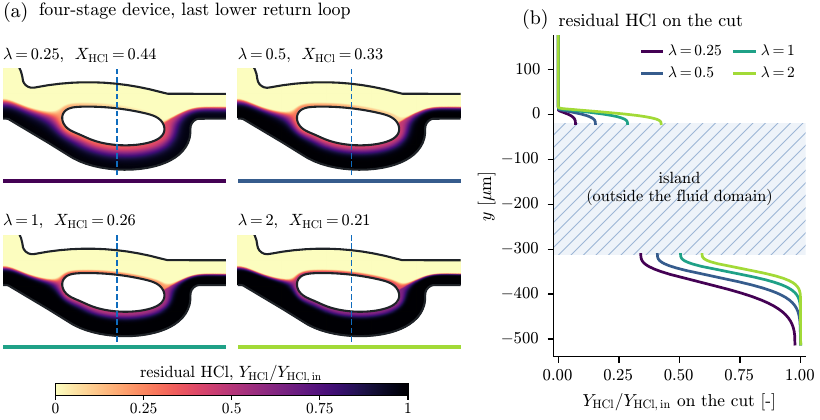}
\caption{{Residual HCl, $Y_{\mathrm{HCl}}/Y_{\mathrm{HCl,in}}$, on the
fine mesh in one four-stage DOE design. (a) Last lower return loop at four
flow rates; dashed line, the cut of (b). (b) Profiles along the cut.}}
\label{fig:lambda1-fields}
\end{figure}

{The residual-HCl fields show this contact time at work
(\cref{fig:lambda1-fields}). HCl is consumed in a thin layer along the
interface with the NaOH stream, which in the last lower return loop wraps
around the island. The flow pattern barely changes with the flow rate, but a
faster stream leaves before this layer has grown, so on a cut through the
loop the layer is thinner and more of the HCl stream passes unreacted
(\cref{fig:lambda1-fields}b). More stages act the other way, giving the layer
a longer path to grow along (\cref{fig:common-resolution-map}b). A lower flow
rate and a longer device therefore act through the same quantity, the
contact time, which is the variable that the equations of
\cref{subsec:results-complexity} find.}

\FloatBarrier

\begin{figure}[H]
\centering
\includegraphics[width=\textwidth]{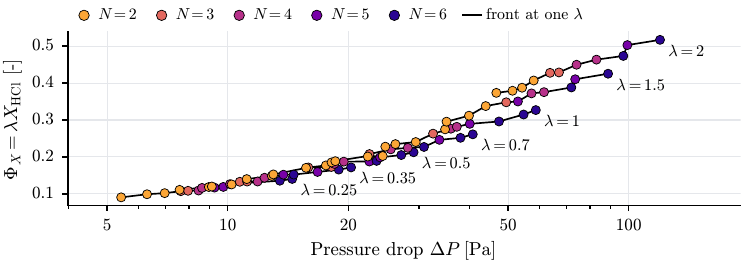}
\caption{{Converted HCl flow $\Phi_X=\lambda X_{\mathrm{HCl}}$ against
pressure drop on the observed fronts of \cref{fig:common-resolution-map}a;
colours as there, flow rate written at the end of each front.}}
\label{fig:converted-flow}
\end{figure}

{The ranking reverses when the amount converted per unit time is
considered. A faster stream brings more HCl, so the converted HCl flow,
$\Phi_X=\lambda X_{\mathrm{HCl}}$ (\cref{eq:converted-throughput}), rises
from 0.14 at $\lambda=0.25$ to 0.52 at $\lambda=2$
(\cref{fig:converted-flow}) while single-pass conversion falls: a low flow
rate converts a larger fraction of less HCl. Plotted against pressure drop,
the fronts of all flow rates fall close to one curve: along the optimised
fronts, the converted flow delivered by one
channel depends mainly on the pressure drop spent, whatever combination of
flow rate and geometry spends it.}

\FloatBarrier

\subsection{{Design equations}}
\label{subsec:results-complexity}

{\Cref{tab:sr-ladder} lists the expressions that the symbolic search
returned as the allowed length grows. Up to 13 nodes for pressure drop and 12
for $Da_{\mathrm{eff}}$, the three seeds returned the same expressions; beyond
that, each seed returns a different one, so the short equations are results
of the search and the long ones are fits. Both responses take the variables in
the same order: flow rate, stage count, and pitch, and only then the return
overhang, loop height, and skew.}

{The first expressions describe a fixed device: pressure drop
proportional to $\lambda$, as in nearly creeping flow, and
$Da_{\mathrm{eff}}\propto\lambda^{-0.43}$. This exponent is close to the
$-1/2$ of two streams that react by diffusion across their interface and far
from the $-1$ of a kinetically controlled reaction (\cref{eq:effective-da}).}

{This exponent is still limited by the mesh:
the reacted layer is thinnest, and least
resolved, at the highest flow rates, and the mean exponent of the eight
geometries simulated on three meshes moves from $-0.20$ on the coarsest to
$-0.43$ on the finest, so the value found here is a lower bound in
magnitude.}

\begin{table}[H]
\centering
\caption{{Expressions returned by the symbolic search as the allowed
length grows (all seven flow rates). Error against the 130 fine-mesh CFD
cases: mean absolute percentage error for $\Delta P$, and mean absolute error
of $X_{\mathrm{HCl}}$ in percentage points (pp) for $Da_{\mathrm{eff}}$. Rows
above ``different in each seed'' were returned identically by the three
seeds; bold, working equations (\cref{subsec:results-symbolic}).}}
\label{tab:sr-ladder}
\small
\renewcommand{\arraystretch}{1.2}
\begin{tabular*}{\textwidth}{@{\extracolsep{\fill}}clll@{}}
\toprule
Nodes & Expression & Enters & Error \\
\midrule
\multicolumn{4}{@{}l}{\emph{Pressure drop,} $\Delta P/\si{Pa}$} \\
3  & $32.6\,\lambda$ & $\lambda$ & 21.8\% \\
5  & $9.50\,\lambda N$ & $N$ & 17.2\% \\
7  & $17.4\,\lambda N^{0.55}$ & & 8.9\% \\
9  & $\lambda\left[14.3+N\,(p/w+4.34)\right]$ & $p/w$ & 5.0\% \\
\textbf{11} & $\lambda\left[17.6+(p/w+4.32)(N-\eta_o)\right]$ & $\eta_o$ & \textbf{4.1\%} \\
13 & $\lambda\left[17.4+(\eta_h+p/w+1.44)(N-\eta_o)\right]$ & $\eta_h$ & 3.0\% \\
17--30 & different in each seed & $\beta$ & 1.5--2.4\% \\
\midrule
\multicolumn{4}{@{}l}{\emph{Effective Damk\"ohler number,} $Da_{\mathrm{eff}}$} \\
5  & $0.297\,\lambda^{-0.43}$ & $\lambda$ & 3.3 pp \\
7  & $0.182\,(N/\lambda)^{0.40}$ & $N$ & 2.3 pp \\
9  & $0.154\,\left[(N+p/w)/\lambda\right]^{0.42}$ & $p/w$ & 1.25 pp \\
\textbf{12} & $0.0606\exp\{[(N+p/w+1.22)/\lambda]^{0.252}\}$ & & \textbf{1.12 pp} \\
14--30 & different in each seed & $\eta_o$, $\eta_h$, $\beta$ & 0.8--1.1 pp \\
\bottomrule
\end{tabular*}
\end{table}

\begin{figure}[H]
\centering
\includegraphics[width=0.58\textwidth]{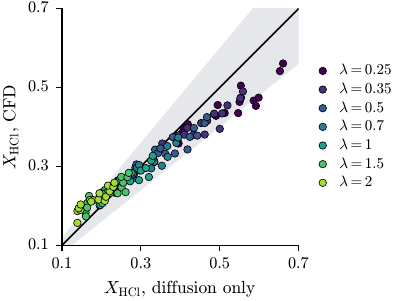}
\caption{{HCl conversion of the 130 fine-mesh cases against a
diffusion-only estimate, coloured by flow rate; line, parity; band,
$\pm20\%$.}}
\label{fig:diffusion-estimate}
\end{figure}

{The reproducible expressions already carry the physics. At nine nodes,
pressure drop is a hydraulic resistance, a fixed loss plus a loss per stage
that grows with pitch, and $Da_{\mathrm{eff}}$ depends on the geometry only
through $N+p/w$, a channel length, which divided by $\lambda$ is a contact
time. A diffusion-only estimate tests this reading: a straight channel in
which the two streams flow side by side at the mean velocity, over the length
of the main centreline, and react where they meet reproduces the conversion
of the 130 fine-mesh cases with a correlation of 0.98 in logarithms
(\cref{fig:diffusion-estimate}). The deviations follow the flow rate: the CFD
lies about 13\% below the estimate at $\lambda=0.25$, where the reacted layer
fills much of the HCl stream, and about 19\% above it at $\lambda=2$, where
the mesh resolves the layer least. At these low Reynolds numbers the stages
therefore act mainly through the contact time they provide, not through the
splitting and recirculation of the Tesla loops, and the loop height, return
overhang, and skew, which change either response by 2--11\% over their
ranges, are secondary levers.}

\FloatBarrier

\subsection{{The front at a flow rate that was not optimised}}
\label{subsec:results-symbolic}

{The equations of \cref{tab:sr-ladder} were fitted to all seven flow
rates. To test whether they carry over to a flow rate they have not seen, the
search is repeated with one flow rate withheld (\cref{subsec:symbolic-method}),
and the predicted fronts are scored by the campaign surrogate of the withheld
flow rate, which on the 74 designs recomputed on the fine mesh predicted
pressure drop within 0.65\% and conversion within 0.78 percentage points on
average.}

\begin{figure}[H]
\centering
\includegraphics[width=\textwidth]{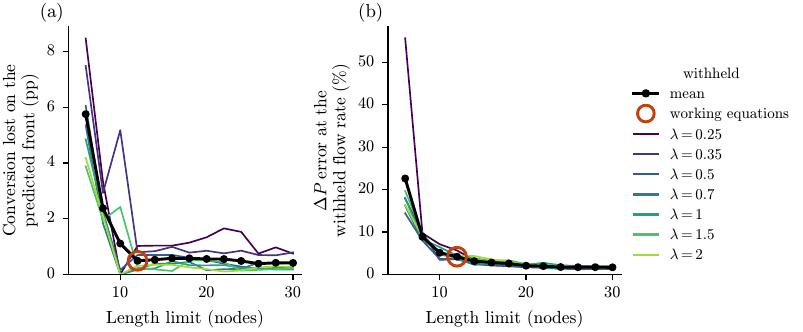}
\caption{{Equations found with one flow rate withheld, against the
length allowed to them. (a) Conversion lost at equal pressure drop by the
predicted front. (b) Mean absolute percentage error of pressure drop against
the fine-mesh CFD of the withheld flow rate. Thin lines, each withheld flow
rate; black line, their mean; circle, working equations.}}
\label{fig:sr-complexity}
\end{figure}

{The first question is how long the equations need to be
(\cref{fig:sr-complexity}). With six nodes, the front predicted at a withheld
flow rate loses 5.8 percentage points of conversion; with 12 nodes it loses
0.49, and longer equations do no better
(\cref{fig:sr-complexity}a). The pressure-drop error at the withheld flow
rate falls in the same way, from 23\% to about 4\%, and the slow gain
beyond does not move the front (\cref{fig:sr-complexity}b). By the rule of
\cref{subsec:symbolic-method}, the working equations are therefore the
12-node expressions, in bold in \cref{tab:sr-ladder}:}
\begin{align}
 \frac{\Delta P}{\SI{1}{Pa}}
 &=\lambda\left[17.6+\left(\frac{p}{w}+4.32\right)\left(N-\eta_o\right)\right],
 \label{eq:pressure-symbolic}\\
 Da_{\mathrm{eff}}
 &=0.0606\,\exp\!\left[\left(\frac{N+p/w+1.22}{\lambda}\right)^{0.252}\right],
 \qquad X_{\mathrm{HCl}}=1-\exp(-Da_{\mathrm{eff}}).
 \label{eq:daeff-symbolic}
\end{align}
{Their conversion error against the 130 fine-mesh cases, 1.12 percentage
points, is of the size of the residual mesh sensitivity of the CFD
(\cref{fig:mesh_response}).}

{Found without one flow rate, the equations still predict the fine-mesh
CFD cases at that flow rate within 4.2\% in pressure drop and 1.2 percentage
points in conversion on average, close to the accuracy of the working
equations fitted to all flow rates. The pressure equation was recovered in
the same form every time, with its three constants within 2\% of those of
\cref{eq:pressure-symbolic}; the conversion equation changed form from one
test to another but always depended on $\lambda$ and $N+p/w$ only.}

\begin{figure}[H]
\centering
\includegraphics[width=0.94\textwidth]{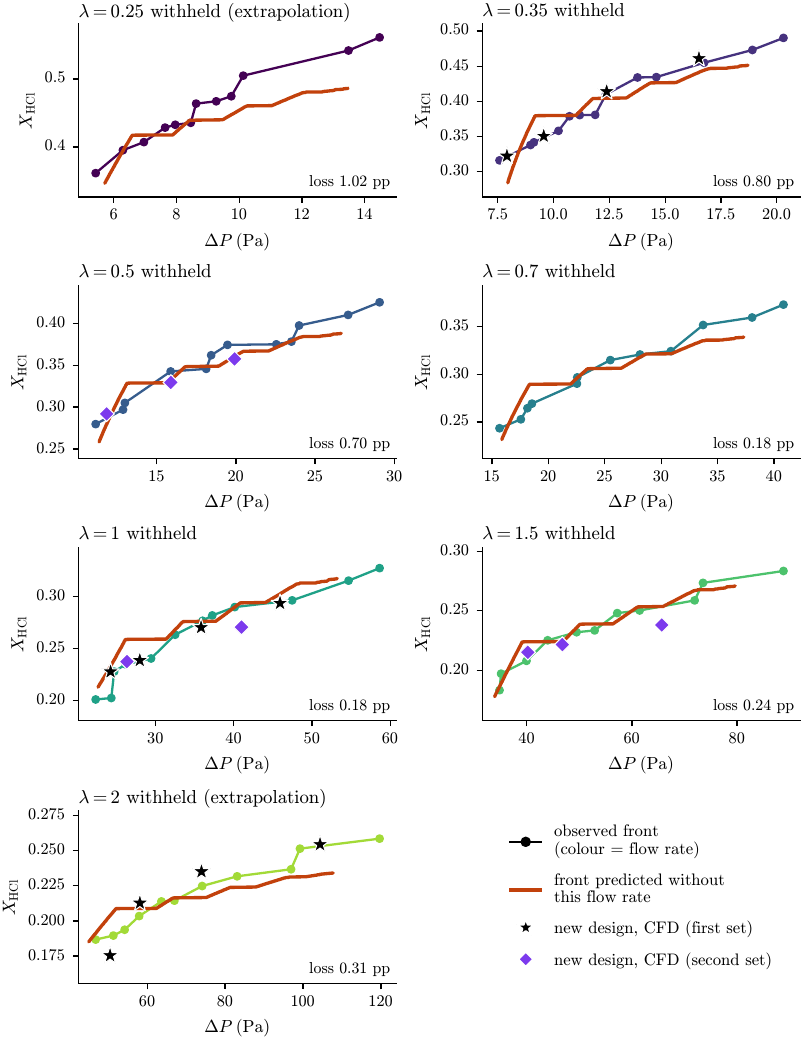}
\caption{{Front at a flow rate withheld from the symbolic search.
Coloured line, nondominated fine-mesh CFD cases from the optimisation at that
flow rate; orange line, front predicted by the 12-node equations found without
it; markers, new designs taken from predicted fronts, at their fine-mesh CFD
values. Loss, conversion lost at equal pressure drop by the predicted front.}}
\label{fig:front-transfer}
\end{figure}

{The fronts read from these equations follow the observed fronts
(\cref{fig:front-transfer}). Judged by the surrogate of the withheld flow
rate, they lose on average 0.49 percentage points of conversion at equal
pressure drop and keep 95\% of the hypervolume. The largest loss, 1.0 point,
is at $\lambda=0.25$, where the prediction extrapolates and falls below the
high-conversion end of the front. The predicted fronts also stop before the
highest-pressure designs, because the search is confined to the region
supported by the six remaining surrogates, and each of their steps is one
stage count.}

{Such a small loss, from equations that ignore two design variables and
never saw the flow rate in question, means that the efficient geometries are
nearly the same at every flow rate. The equations predict this: pressure drop
per unit flow rate and the length $N+p/w$ depend on the geometry alone, so the
geometries that give the most length for a given pressure drop per unit flow
rate are the same at every flow rate. What moves with flow rate is the
position of the front, and the equations describe that movement.}

{Short equations also choose better than a flexible model. A single
Gaussian process trained on the CFD cases of the other six campaigns, with
$\lambda$ as an input, predicts the withheld CFD cases more closely than
\cref{eq:pressure-symbolic,eq:daeff-symbolic} but selects a worse front,
losing 1.7 percentage points, because the optimiser seeks out the designs
whose conversion a flexible model overestimates.}

{Finally, twenty designs taken from predicted fronts were simulated on
the fine mesh: twelve from the fronts of a preliminary fit of fixed power-law
form, and eight read from the fronts of
\cref{eq:pressure-symbolic,eq:daeff-symbolic}. All met the acceptance criteria
of \cref{subsec:fidelity}, and the equations predicted them within 4.2\% in
pressure drop and 1.3 percentage points in conversion, the same accuracy as
for the cases produced by the optimisation. Relative to the front already
observed at their flow rate, the first twelve lie on average 0.9 percentage
points above it and the last eight 0.5 points below it, the loss that the
withheld-flow-rate test had estimated. The largest shortfall, 1.9 points, is
within the change in conversion under further mesh refinement
(\cref{fig:mesh_response}).}

{The last eight fall short because of a variable the short equations leave
out. \Cref{eq:pressure-symbolic} lowers the pressure drop as the return
overhang grows, while \cref{eq:daeff-symbolic} does not contain $\eta_o$, so
every design read from the two equations has $\eta_o$ near its upper bound.
The conversion paid for that choice is absent from the short equation, which
overestimated conversion in seven of these eight designs, by 1.1 percentage
points on average; equations long enough to contain $\eta_o$ (16 nodes or
more, fitted to all flow rates and therefore not a blind test) remove the
bias. The 12-node equations therefore place the front of an unseen flow rate
within about one percentage point of conversion; the last point still
requires the optimisation, or a fine-mesh check of a few candidates with
$\eta_o$ varied.}

{This accuracy comes at a small fraction of the cost of an optimisation
(\cref{tab:cost}). At a flow rate inside the tested range, the front read
from the equations costs no CFD, and a fine-mesh check of four candidates on
it costs about two hours, against about 41 hours for a new campaign and the
fine-mesh refinement of its front. The saving holds only inside the flow range
and design space already sampled, and only once campaigns exist to fit the
equations to; the withheld-flow-rate test shows that six of them suffice to
predict the seventh.}

\begin{table}[H]
\centering
\caption{{CFD cost of the front at one flow rate inside the tested range.
Times are CFD wall times summed over the runs, mean over the seven flow
rates.}}
\label{tab:cost}
\small
\begin{tabular*}{\textwidth}{@{\extracolsep{\fill}}lrr@{}}
\toprule
Route to the front & CFD runs & CFD time (h) \\
\midrule
New optimisation campaign (\cref{fig:hv-stages}) & 153 & 34 \\
\quad plus fine-mesh refinement of its front & 10 & 7 \\
Front read from \cref{eq:pressure-symbolic,eq:daeff-symbolic} & 0 & -- \\
\quad plus fine-mesh check of four candidates & 4 & 2 \\
\bottomrule
\end{tabular*}
\end{table}

{Once tested, the two working equations also put a price on each lever.
One more stage adds 15--18\% to the pressure drop and 6--7\% to
$Da_{\mathrm{eff}}$; halving the flow rate through the same device halves the
pressure drop and raises $Da_{\mathrm{eff}}$ by at least a third. Raising the
flow rate per channel is the opposite move: keeping conversion while the flow
rate doubles requires $N+p/w+1.22$ to double, and \cref{eq:pressure-symbolic}
then gives about four times the pressure drop, whatever the converged
exponent of $\lambda$; from three stages at $p/w=1.5$, about nine stages are
needed, beyond the six tested. If conversion may fall instead, the converted
flow of the 83 nondominated fine-mesh designs follows
$\Phi_X\approx0.031\,(\Delta P/\SI{1}{Pa})^{0.61}$ ($R^2=0.98$ in
logarithms; 0.59 from the equations), and doubling it costs about three times
the pressure drop.}

\FloatBarrier

\section{Conclusions}
\label{sec:conclusion}

{Short equations found by symbolic regression of the optimisation
surrogates recover the pressure--conversion front at a flow rate that was not
optimised. Found without that flow rate, the 12-node equations placed its
front within about half a percentage point of conversion of the optimised
one, less than half the spacing between neighbouring designs on a front and
under 4\% of the conversion range it spans. New fine-mesh designs read from
these fronts confirmed both the accuracy of the equations and the size of
that shortfall. Once campaigns at a few flow rates exist, and six were enough
here to predict the seventh, a new flow rate inside the tested range costs a
fine-mesh check of a few designs read from the equations, about two hours of
CFD, instead of a new optimisation of about 40 hours.}

{The front transfers because the equations are short and their form
follows the physics. Given no functional form, three independent searches
returned the same equations: pressure drop is a hydraulic resistance
proportional to flow rate, and the effective Damk\"ohler number is set by a
channel length divided by flow rate, a contact time. A diffusion-only estimate
reproduces the fine-mesh conversion, so in this two-dimensional laminar model
the stages act through contact time rather than through added mixing. Both
equations depend on the geometry only through quantities that do not involve
the flow rate, so the efficient geometries change little with flow rate: the
front moves, and the equations describe how. Longer equations fitted the data
better without predicting better fronts.}

For the Tesla-valve family itself, the principal engineering outcome is
{the price of throughput. Over the eightfold flow range, the highest
single-pass conversion fell by half, while the converted HCl flow per channel
rose almost fourfold.} Thus, scale-up does not simply
make a design better or worse: it exchanges conversion per pass for a larger
converted HCl flow while increasing the pressure penalty. {The equations
turn this into a rule: keeping conversion while doubling the flow rate per
channel requires twice the length and about four times the pressure drop.
Stage count and pitch are the levers that matter; loop height, return
overhang, and skew are secondary, yet still decide the last point of a
front.}

{These results hold within the limits of the study. The CFD model is
two-dimensional, steady, and laminar, and has not been validated
experimentally. Conversion keeps a residual mesh sensitivity that grows with
flow rate, so margins below about two percentage points at the highest flow
rates are not certified; finer meshes would lower conversion most there and so
raise, not lower, the price of throughput. The equations were tested only
inside the flow range and design space sampled, and give no probability of
meeting a constraint, so a hard constraint still requires fine-mesh
confirmation. The last point of a front, decided by variables that the short
equations omit, still requires the optimisation or a fine-mesh check of a few
candidates. The procedure itself is not specific to the Tesla valve: it
needs optimisation campaigns at several values of an operating variable. In
three-dimensional channels and at higher Reynolds numbers, where the Tesla
loops can stretch the interface between the streams, conversion above the
diffusion-only estimate would measure the mixing that the stages add, and
would show whether equations this short still transfer the front.}
\label{subsec:results-limitations}

\section*{Data availability}
The data supporting this study are provided as supplementary data, in CSV
files with a description of each column: every evaluation of the seven
optimisation campaigns, the 130 fine-mesh cases with their diffusion-only
estimate, the mesh-independence study, the twenty new designs, the scores of
the withheld-flow-rate test, and the equations returned by the symbolic
search. The Supplementary Material gives the geometry definition, the solver
and mesh settings with their acceptance criteria, the optimisation protocol,
and the symbolic-regression settings needed to reproduce the simulations and
the analysis. Solver case and data files are available from the corresponding
author on reasonable request.

\section*{Acknowledgements}
The authors gratefully acknowledge financial support from the Water4All
project.

\section*{Declaration of generative AI and AI-assisted technologies in the
manuscript preparation process}
During the preparation of this work, the authors used OpenAI Codex and Anthropic Claude to support
manuscript organisation, language editing, consistency checks, and preparation
of submission files. The authors reviewed and edited the output as needed and
take full responsibility for the content of the published article.

\bibliography{refs_tesla}

\begin{thebibliography}{10}
\expandafter\ifx\csname url\endcsname\relax
  \def\url#1{\texttt{#1}}\fi
\expandafter\ifx\csname urlprefix\endcsname\relax\def\urlprefix{URL }\fi
\expandafter\ifx\csname href\endcsname\relax
  \def\href#1#2{#2} \def\path#1{#1}\fi

\bibitem{Cai2023}
S.~Cai, Y.~Jin, Y.~Lin, Y.~He, P.~Zhang, Z.~Ge, W.~Yang, Micromixing within
  microfluidic devices: Fundamentals, design, and fabrication, Biomicrofluidics
  17~(6) (2023) 061503.
\newblock \href {https://doi.org/10.1063/5.0178396}
  {\path{doi:10.1063/5.0178396}}.

\bibitem{Raza2020}
W.~Raza, S.~Hossain, K.-Y. Kim, A review of passive micromixers with a
  comparative analysis, Micromachines 11~(5) (2020) 455.
\newblock \href {https://doi.org/10.3390/mi11050455}
  {\path{doi:10.3390/mi11050455}}.

\bibitem{Ganguli2024}
A.~Ganguli, V.~Bhatt, A.~Yagodnitsyna, D.~Pinjari, A.~Pandit, A review of
  pressure drop and mixing characteristics in passive mixers involving miscible
  liquids, Micromachines 15~(6) (2024) 691.
\newblock \href {https://doi.org/10.3390/mi15060691}
  {\path{doi:10.3390/mi15060691}}.

\bibitem{Su2014}
Y.~Su, A.~Lautenschleger, G.~Chen, E.~Y. Kenig, A numerical study on liquid
  mixing in multichannel micromixers, Industrial \& Engineering Chemistry
  Research 53~(1) (2014) 390--401.
\newblock \href {https://doi.org/10.1021/ie401924x}
  {\path{doi:10.1021/ie401924x}}.

\bibitem{Nogueira2019SPSO}
I.~B.~R. Nogueira, M.~A.~F. Martins, R.~Requi{\~a}o, A.~R. Oliveira, V.~Viena,
  H.~Koivisto, A.~E. Rodrigues, J.~M. Loureiro, A.~M. Ribeiro, Optimization of
  a true moving bed unit and determination of its feasible operating region
  using a novel sliding particle swarm optimization, Computers \& Industrial
  Engineering 135 (2019) 368--381.
\newblock \href {https://doi.org/10.1016/j.cie.2019.06.020}
  {\path{doi:10.1016/j.cie.2019.06.020}}.

\bibitem{Nogueira2022NestedLoop}
I.~B.~R. Nogueira, R.~O.~M. Dias, C.~M. Rebello, E.~A. Costa, V.~V. Santana,
  A.~E. Rodrigues, A.~Ferreira, A.~M. Ribeiro, A novel nested loop optimization
  problem based on deep neural networks and feasible operation regions
  definition for simultaneous material screening and process optimization,
  Chemical Engineering Research and Design 180 (2022) 243--253.
\newblock \href {https://doi.org/10.1016/j.cherd.2022.02.013}
  {\path{doi:10.1016/j.cherd.2022.02.013}}.

\bibitem{Dong2021}
Z.~Dong, Z.~Wen, F.~Zhao, S.~Kuhn, T.~No{\"e}l, Scale-up of micro- and
  milli-reactors: An overview of strategies, design principles and
  applications, Chemical Engineering Science: X 10 (2021) 100097.
\newblock \href {https://doi.org/10.1016/j.cesx.2021.100097}
  {\path{doi:10.1016/j.cesx.2021.100097}}.

\bibitem{Abolpour2022}
B.~Abolpour, R.~Hekmatkhah, R.~Shamsoddini, Optimum design for the tesla
  micromixer, Microfluidics and Nanofluidics 26~(6) (2022) 46.
\newblock \href {https://doi.org/10.1007/s10404-022-02530-6}
  {\path{doi:10.1007/s10404-022-02530-6}}.

\bibitem{Guo2022}
K.~Guo, Y.~Chen, Z.~Zhou, S.~Zhu, Z.~Ni, N.~Xiang, A novel 3d tesla valve
  micromixer for efficient mixing and chitosan nanoparticle production,
  Electrophoresis 43~(21--22) (2022) 2184--2194.
\newblock \href {https://doi.org/10.1002/elps.202200077}
  {\path{doi:10.1002/elps.202200077}}.

\bibitem{Sun2024}
D.~Sun, L.~Zeng, Y.~Yang, C.~Liu, J.~Hong, W.~Han, W.~Li, C.~Wang, J.~Shen,
  H.~Yang, H.~Zhang, Numerical and experimental investigation of tesla
  micromixers with different three-dimensional herringbone structures, Chemical
  Engineering and Processing - Process Intensification 205 (2024) 110040.
\newblock \href {https://doi.org/10.1016/j.cep.2024.110040}
  {\path{doi:10.1016/j.cep.2024.110040}}.

\bibitem{Li2025}
J.~Li, B.~Ma, X.~Zhang, Q.~Zhao, R.~Sun, J.~Zhao, Design and optimization of
  tesla micromixer with asymmetrical arrangement for efficient mixing in
  microfluidic chip, Chemical Engineering and Processing - Process
  Intensification 209 (2025) 110181.
\newblock \href {https://doi.org/10.1016/j.cep.2025.110181}
  {\path{doi:10.1016/j.cep.2025.110181}}.

\bibitem{Liu2025}
P.~Liu, Y.~Yue, Y.~Guo, W.~Yang, L.~Zang, Baffle optimization design and mixing
  performance study of asymmetric tesla vortex micromixer, International
  Communications in Heat and Mass Transfer 169 (2025) 109663.
\newblock \href {https://doi.org/10.1016/j.icheatmasstransfer.2025.109663}
  {\path{doi:10.1016/j.icheatmasstransfer.2025.109663}}.

\bibitem{Yang2025}
Y.~Yang, K.~Cui, G.~Song, J.~Shi, Optimal design of passive tesla micromixers
  for enhanced mixing and gradient performance in multiphase material forming,
  Journal of Applied Physics 138~(5) (2025) 054702.
\newblock \href {https://doi.org/10.1063/5.0287223}
  {\path{doi:10.1063/5.0287223}}.

\bibitem{Juraeva2022Tesla}
M.~Juraeva, D.-J. Kang, Mixing performance of the modified {Tesla} micromixer
  with tip clearance, Micromachines 13~(9) (2022) 1375.
\newblock \href {https://doi.org/10.3390/mi13091375}
  {\path{doi:10.3390/mi13091375}}.

\bibitem{WangChen2022SA}
H.~Wang, X.~Chen, Optimization of {T}esla valve-typed micromixer based on
  simulated annealing algorithm, Surface Review and Letters 29~(7) (2022)
  2250094.
\newblock \href {https://doi.org/10.1142/S0218625X22500949}
  {\path{doi:10.1142/S0218625X22500949}}.

\bibitem{Xiong2023}
M.~Xiong, J.~Yang, X.~Ding, H.~Li, H.~Zhang, Topology optimization design of
  micromixer based on principle of tesla valve: An experimental and numerical
  study, Chemical Engineering and Processing - Process Intensification 193
  (2023) 109560.
\newblock \href {https://doi.org/10.1016/j.cep.2023.109560}
  {\path{doi:10.1016/j.cep.2023.109560}}.

\bibitem{LiTesla2026}
S.~Li, Y.~Chen, M.~Li, X.~Xu, C.~Wu, G.~Yang, Simulation and experimental study
  of the micromixer with tesla valve structure, Chemical Engineering and
  Processing - Process Intensification 227 (2026) 110892.
\newblock \href {https://doi.org/10.1016/j.cep.2026.110892}
  {\path{doi:10.1016/j.cep.2026.110892}}.

\bibitem{Raza2018}
W.~Raza, S.-B. Ma, K.-Y. Kim, Multi-objective optimizations of a serpentine
  micromixer with crossing channels at low and high {R}eynolds numbers,
  Micromachines 9~(3) (2018) 110.
\newblock \href {https://doi.org/10.3390/mi9030110}
  {\path{doi:10.3390/mi9030110}}.

\bibitem{Fonseca2025BayesShape}
R.~Fonseca, F.~Bernardo, Design of microfluidic mixers using {Bayesian} shape
  optimization, Systems and Control Transactions 4 (2025) 1017--1022.
\newblock \href {https://doi.org/10.69997/sct.199876}
  {\path{doi:10.69997/sct.199876}}.

\bibitem{Baldyga1999}
J.~Ba{\l}dyga, J.~R. Bourne, Turbulent Mixing and Chemical Reactions, Wiley,
  Chichester, 1999.

\bibitem{Bourne2003}
J.~R. Bourne, Mixing and the selectivity of chemical reactions, Organic Process
  Research \& Development 7~(4) (2003) 471--508.
\newblock \href {https://doi.org/10.1021/op020074q}
  {\path{doi:10.1021/op020074q}}.

\bibitem{Lima2026FuzzyCFD}
P.~Lima, L.~S. de~Souza, L.~Schnitman, G.~D. Reges, T.~P.~G. Mendes, E.~G.~S.
  Nascimento, I.~Nogueira, Fuzzy identification of transitional flow patterns
  in core annular multiphase flows using computational fluid dynamics, Chemical
  Engineering Research and Design 228 (2026) 150--171.
\newblock \href {https://doi.org/10.1016/j.cherd.2026.02.048}
  {\path{doi:10.1016/j.cherd.2026.02.048}}.

\bibitem{Lima2025CAFControl}
P.~Lima, E.~Costa, T.~P.~G. Mendes, L.~Schnitman, S.~Skogestad, I.~Nogueira,
  Simple control structure for stabilizing core annular flow operation in heavy
  oil transportation, Computers \& Chemical Engineering 194 (2025) 108978.
\newblock \href {https://doi.org/10.1016/j.compchemeng.2024.108978}
  {\path{doi:10.1016/j.compchemeng.2024.108978}}.

\bibitem{Shahriari2016}
B.~Shahriari, K.~Swersky, Z.~Wang, R.~P. Adams, N.~de~Freitas, Taking the human
  out of the loop: A review of bayesian optimization, Proceedings of the IEEE
  104~(1) (2016) 148--175.
\newblock \href {https://doi.org/10.1109/JPROC.2015.2494218}
  {\path{doi:10.1109/JPROC.2015.2494218}}.

\bibitem{Frazier2018}
P.~I. Frazier, A tutorial on bayesian optimization, arXiv preprint
  arXiv:1807.02811 (2018).
\newblock \href {https://doi.org/10.48550/arXiv.1807.02811}
  {\path{doi:10.48550/arXiv.1807.02811}}.

\bibitem{Garnett2023}
R.~Garnett, Bayesian Optimization, Cambridge University Press, Cambridge, 2023.
\newblock \href {https://doi.org/10.1017/9781108348973}
  {\path{doi:10.1017/9781108348973}}.

\bibitem{Kundacina2025}
I.~Kundacina, O.~Kundacina, D.~Miskovic, V.~Radonic, Advancing microfluidic
  design with machine learning: A bayesian optimization approach, Lab on a Chip
  25~(4) (2025) 657--672.
\newblock \href {https://doi.org/10.1039/D4LC00872C}
  {\path{doi:10.1039/D4LC00872C}}.

\bibitem{Basha2023CEJ}
N.~Basha, T.~Savage, J.~McDonough, E.~A. del R{\'i}o~Chanona, O.~K. Matar,
  Discovery of mixing characteristics for enhancing coiled reactor performance
  through a {Bayesian} optimisation-{CFD} approach, Chemical Engineering
  Journal 473 (2023) 145217.
\newblock \href {https://doi.org/10.1016/j.cej.2023.145217}
  {\path{doi:10.1016/j.cej.2023.145217}}.

\bibitem{Savage2024NatChemEng}
T.~Savage, N.~Basha, J.~McDonough, J.~Krassowski, O.~Matar, E.~A. del
  R{\'i}o~Chanona, Machine learning-assisted discovery of flow reactor designs,
  Nature Chemical Engineering 1~(8) (2024) 522--531.
\newblock \href {https://doi.org/10.1038/s44286-024-00099-1}
  {\path{doi:10.1038/s44286-024-00099-1}}.

\bibitem{Zhao2026}
F.~Zhao, Y.~Huang, C.~Mei, X.~Chen, Design centering via {Bayesian
  Optimization} for robust operation of chemical processes, Industrial \&
  Engineering Chemistry Research 65~(16) (2026) 8355--8372.
\newblock \href {https://doi.org/10.1021/acs.iecr.5c04321}
  {\path{doi:10.1021/acs.iecr.5c04321}}.

\bibitem{Lima2026Burner}
P.~S. Lima, P.~R.~S. dos Reis, A.~{\'A}.~B. Santos, E.~A. del R{\'i}o~Chanona,
  I.~B.~R. Nogueira, Multiobjective and multifidelity bayesian optimization
  {CFD}-based design and operation optimization of non-premixed hydrogen
  burners, Computers \& Chemical Engineering 214 (2026) 109787.
\newblock \href {https://doi.org/10.1016/j.compchemeng.2026.109787}
  {\path{doi:10.1016/j.compchemeng.2026.109787}}.

\bibitem{Savage2023}
T.~Savage, N.~Basha, J.~McDonough, O.~K. Matar, E.~A. del R{\'i}o~Chanona,
  Multi-fidelity data-driven design and analysis of reactor and tube
  simulations, Computers \& Chemical Engineering 179 (2023) 108410.
\newblock \href {https://doi.org/10.1016/j.compchemeng.2023.108410}
  {\path{doi:10.1016/j.compchemeng.2023.108410}}.

\bibitem{Lima2026BurnerMFBO}
P.~S. Lima, P.~R.~S. dos Reis, A.~{\'A}.~B. Santos, E.~A. del R{\'i}o~Chanona,
  I.~B.~R. Nogueira, Multi-fidelity bayesian optimization framework for
  {CFD}-based non-premixed burner design, Results in Engineering 32 (2026)
  112628.
\newblock \href {https://doi.org/10.1016/j.rineng.2026.112628}
  {\path{doi:10.1016/j.rineng.2026.112628}}.

\bibitem{Swersky2013}
K.~Swersky, J.~Snoek, R.~P. Adams,
  \href{https://proceedings.neurips.cc/paper_files/paper/2013/hash/f33ba15effa5c10e873bf3842afb46a6-Abstract.html}{Multi-task
  {B}ayesian optimization}, in: Advances in Neural Information Processing
  Systems 26, 2013, pp. 2004--2012.
\newline\urlprefix\url{https://proceedings.neurips.cc/paper_files/paper/2013/hash/f33ba15effa5c10e873bf3842afb46a6-Abstract.html}

\bibitem{Krause2011}
A.~Krause, C.~S. Ong,
  \href{https://proceedings.neurips.cc/paper_files/paper/2011/hash/f3f1b7fc5a8779a9e618e1f23a7b7860-Abstract.html}{Contextual
  {G}aussian process bandit optimization}, in: Advances in Neural Information
  Processing Systems 24, 2011, pp. 2447--2455.
\newline\urlprefix\url{https://proceedings.neurips.cc/paper_files/paper/2011/hash/f3f1b7fc5a8779a9e618e1f23a7b7860-Abstract.html}

\bibitem{WeaverRosen2023}
J.~M. Weaver-Rosen, R.~J. Malak, An algorithm for multi-objective efficient
  parametric optimization, Journal of Mechanical Design 145~(3) (2023) 031709.
\newblock \href {https://doi.org/10.1115/1.4056764}
  {\path{doi:10.1115/1.4056764}}.

\bibitem{Cheng2026PPSLMOBO}
J.~Cheng, B.~Xue, Q.~Zhang, Parametric {P}areto set learning for expensive
  multi-objective optimization, Proceedings of the AAAI Conference on
  Artificial Intelligence 40~(43) (2026) 36829--36837.
\newblock \href {https://doi.org/10.1609/aaai.v40i43.41009}
  {\path{doi:10.1609/aaai.v40i43.41009}}.

\bibitem{Oberdieck2016}
R.~Oberdieck, N.~A. Diangelakis, I.~Nascu, M.~M. Papathanasiou, M.~Sun,
  S.~Avraamidou, E.~N. Pistikopoulos, On multi-parametric programming and its
  applications in process systems engineering, Chemical Engineering Research
  and Design 116 (2016) 61--82.
\newblock \href {https://doi.org/10.1016/j.cherd.2016.09.034}
  {\path{doi:10.1016/j.cherd.2016.09.034}}.

\bibitem{Cozad2014}
A.~Cozad, N.~V. Sahinidis, D.~C. Miller, Learning surrogate models for
  simulation-based optimization, AIChE Journal 60~(6) (2014) 2211--2227.
\newblock \href {https://doi.org/10.1002/aic.14418}
  {\path{doi:10.1002/aic.14418}}.

\bibitem{Wilson2017}
Z.~T. Wilson, N.~V. Sahinidis, The {ALAMO} approach to machine learning,
  Computers \& Chemical Engineering 106 (2017) 785--795.
\newblock \href {https://doi.org/10.1016/j.compchemeng.2017.02.010}
  {\path{doi:10.1016/j.compchemeng.2017.02.010}}.

\bibitem{Engle2022}
M.~R. Engle, N.~V. Sahinidis, Deterministic symbolic regression with derivative
  information: General methodology and application to equations of state, AIChE
  Journal 68~(6) (2022) e17457.
\newblock \href {https://doi.org/10.1002/aic.17457}
  {\path{doi:10.1002/aic.17457}}.

\bibitem{Rebello2022MLGuide}
C.~M. Rebello, P.~H. Marrocos, E.~A. Costa, V.~V. Santana, A.~E. Rodrigues,
  A.~M. Ribeiro, I.~B.~R. Nogueira, Machine learning-based dynamic modeling for
  process engineering applications: A guideline for simulation and prediction
  from perceptron to deep learning, Processes 10~(2) (2022) 250.
\newblock \href {https://doi.org/10.3390/pr10020250}
  {\path{doi:10.3390/pr10020250}}.

\bibitem{Nogueira2022UDE}
I.~B.~R. Nogueira, V.~V. Santana, A.~M. Ribeiro, A.~E. Rodrigues, Using
  scientific machine learning to develop universal differential equation for
  multicomponent adsorption separation systems, The Canadian Journal of
  Chemical Engineering 100~(9) (2022) 2279--2290.
\newblock \href {https://doi.org/10.1002/cjce.24495}
  {\path{doi:10.1002/cjce.24495}}.

\bibitem{Lima2023UDECryst}
F.~A. R.~D. Lima, C.~M. Rebello, E.~A. Costa, V.~V. Santana, M.~G.~F.
  de~Moares, A.~G. Barreto, A.~R. Secchi, M.~B. de~Souza, I.~B.~R. Nogueira,
  Improved modeling of crystallization processes by universal differential
  equations, Chemical Engineering Research and Design 200 (2023) 538--549.
\newblock \href {https://doi.org/10.1016/j.cherd.2023.11.032}
  {\path{doi:10.1016/j.cherd.2023.11.032}}.

\bibitem{Ansari2022}
M.~Ansari, H.~A. Gandhi, D.~G. Foster, A.~D. White, Iterative symbolic
  regression for learning transport equations, AIChE Journal 68~(6) (2022)
  e17695.
\newblock \href {https://doi.org/10.1002/aic.17695}
  {\path{doi:10.1002/aic.17695}}.

\bibitem{Rogers2024}
A.~W. Rogers, A.~Lane, C.~Mendoza, S.~Watson, A.~Kowalski, P.~Martin, D.~Zhang,
  Integrating knowledge-guided symbolic regression and model-based design of
  experiments to automate process flow diagram development, Chemical
  Engineering Science 300 (2024) 120580.
\newblock \href {https://doi.org/10.1016/j.ces.2024.120580}
  {\path{doi:10.1016/j.ces.2024.120580}}.

\bibitem{Rebello2024SciML}
C.~M. Rebello, E.~A. Costa, M.~Fontana, L.~Schnitman, I.~B.~R. Nogueira,
  Interpretable scientific machine learning approach for correcting
  phenomenological models: Methodology validation on an {ESP} prototype,
  Industrial \& Engineering Chemistry Research 63~(44) (2024) 19030--19050.
\newblock \href {https://doi.org/10.1021/acs.iecr.4c02104}
  {\path{doi:10.1021/acs.iecr.4c02104}}.

\bibitem{Schmidt2009}
M.~Schmidt, H.~Lipson, Distilling free-form natural laws from experimental
  data, Science 324~(5923) (2009) 81--85.
\newblock \href {https://doi.org/10.1126/science.1165893}
  {\path{doi:10.1126/science.1165893}}.

\bibitem{Cranmer2023PySR}
M.~Cranmer, Interpretable machine learning for science with {PySR} and
  {SymbolicRegression.jl}, arXiv preprint arXiv:2305.01582 (2023).
\newblock \href {https://doi.org/10.48550/arXiv.2305.01582}
  {\path{doi:10.48550/arXiv.2305.01582}}.

\bibitem{Brunton2016}
S.~L. Brunton, J.~L. Proctor, J.~N. Kutz, Discovering governing equations from
  data by sparse identification of nonlinear dynamical systems, Proceedings of
  the National Academy of Sciences 113~(15) (2016) 3932--3937.
\newblock \href {https://doi.org/10.1073/pnas.1517384113}
  {\path{doi:10.1073/pnas.1517384113}}.

\bibitem{Miettinen1999}
K.~Miettinen, Nonlinear Multiobjective Optimization, Kluwer Academic
  Publishers, Boston, 1998.
\newblock \href {https://doi.org/10.1007/978-1-4615-5563-6}
  {\path{doi:10.1007/978-1-4615-5563-6}}.

\bibitem{Geuzaine2009}
C.~Geuzaine, J.-F. Remacle, {Gmsh}: A 3-d finite element mesh generator with
  built-in pre- and post-processing facilities, International Journal for
  Numerical Methods in Engineering 79~(11) (2009) 1309--1331.
\newblock \href {https://doi.org/10.1002/nme.2579}
  {\path{doi:10.1002/nme.2579}}.

\bibitem{McKay1979}
M.~D. McKay, R.~J. Beckman, W.~J. Conover, A comparison of three methods for
  selecting values of input variables in the analysis of output from a computer
  code, Technometrics 21~(2) (1979) 239--245.
\newblock \href {https://doi.org/10.1080/00401706.1979.10489755}
  {\path{doi:10.1080/00401706.1979.10489755}}.

\bibitem{Rasmussen2006}
C.~E. Rasmussen, C.~K.~I. Williams, Gaussian Processes for Machine Learning,
  MIT Press, Cambridge, MA, 2006.

\bibitem{Zitzler1999}
E.~Zitzler, L.~Thiele, Multiobjective evolutionary algorithms: A comparative
  case study and the strength pareto approach, IEEE Transactions on
  Evolutionary Computation 3~(4) (1999) 257--271.
\newblock \href {https://doi.org/10.1109/4235.797969}
  {\path{doi:10.1109/4235.797969}}.

\bibitem{Mikkola2023}
P.~Mikkola, J.~Martinelli, L.~Filstroff, S.~Kaski,
  \href{https://proceedings.mlr.press/v206/mikkola23a.html}{Multi-fidelity
  bayesian optimization with unreliable information sources}, in: Proceedings
  of the 26th International Conference on Artificial Intelligence and
  Statistics, Vol. 206 of Proceedings of Machine Learning Research, PMLR, 2023,
  pp. 7425--7454.
\newline\urlprefix\url{https://proceedings.mlr.press/v206/mikkola23a.html}

\end{thebibliography}

\end{document}